# Roughness of oxide glass sub-critical fracture surfaces

Gaël Pallares[1,2], Frédéric Lechenault[3], Matthieu George[1], Elisabeth Bouchaud[4], Cédric Ottina[2], Cindy L. Rountree[2] and Matteo Ciccotti[5,*]

[1] Laboratoire Charles Coulomb (UMR 5221, Université Montpellier, CNRS) Montpellier, France

[2] SPEC, CEA, CNRS, Université Paris-Saclay, CEA Saclay, Gif-sur-Yvette, France

[3] Laboratoire de Physique Statistique (UMR 8550, Ecole Normale Supérieure, CNRS, UPMC Univ Paris 6, Université Paris Diderot) Paris, France

[4] Laboratoire Gulliver, EC2M (UMR 7083, ESPCI Paris, PSL Research University) Paris, France

[5] Laboratoire de Science et Ingénierie de la Matière Molle (UMR 6715, ESPCI Paris, CNRS, UPMC Univ Paris 6, PSL Research University) Paris, France

This work was supported by the ANR Grant "Corcosil" No. ANR-07-BLAN-0261-02.

[*] Author to whom correspondence should be addressed. e-mail: matteo.ciccotti@espci.fr

Accepted for publication on Journal of the American Ceramic Society on September 26th 2017

## Abstract

An original setup combining a very stable loading stage, an atomic force microscope and an environmental chamber, allows to obtain very stable sub-critical fracture propagation in oxide glasses under controlled environment, and subsequently to finely characterize the nanometric roughness properties of the crack surfaces. The analysis of the surface roughness is conducted both in terms of the classical root mean square roughness to compare with the literature, and in terms of more physically adequate indicators related to the self-affine nature of the fracture surfaces. Due to the comparable nanometric scale of the surface roughness, the AFM tip size and the instrumental noise, a special care is devoted to the statistical evaluation of the metrologic properties. The roughness amplitude of several oxide glasses was shown to decrease as a function of the stress intensity factor, to be quite insensitive to the relative humidity and to increase with the degree of heterogeneity of the glass. The results are discussed in terms of several modeling arguments concerning the coupling between crack propagation, material's heterogeneity, crack tip plastic deformation and water diffusion at the crack tip. A synthetic new model is presented combining the predictions of a model by Wiederhorn et al. [1] on the effect of the material's heterogeneity on the crack tip stresses with the self-affine nature of the fracture surfaces.

# 1 Introduction

Fractography has been widely used for post-mortem investigation of the failure history and of the cause of its initiation, and it has become an invaluable tool for understanding the crack propagation mechanisms in different materials [2, 3]. For very brittle materials, such as glasses or ceramics, the fracture surface is classically divided into mirror, mist and hackle zones, depending on the appearance they present when observed by optical microscopy. These three regions are always present close to the site of initiation of an unstable crack and they form approximately concentric regions, the size of which can be empirically related to the stress conditions at the moment of failure [4]. This feature is originated by a progressive roughening of the fracture surfaces as the crack keeps accelerating during unstable fracture. It was attributed either to the enhanced damage induced by the increase of stress intensity factor $K$, or to the dynamic effects related to the increase of the fracture velocity, or to the fracture front fragmentation induced by mode III twist [5, 2, 6].

The development of local-probe microscopy in the 90's has shed new light on the nature of the mirror region, revealing more rich fracture roughness patterns where the RMS amplitude can be as small as a fraction of nanometer (for micron-size images) [7, 8, 9]. The fracture surface roughening during unstable crack acceleration was thus shown to begin well before the end of the optically flat mirror region [5]. Moreover, the same kind of roughening pattern was observed to occur at very small lateral scales around the initiation site in high strength optical fibers [2]. It is presently recognized that the size of the mirror region cannot be simply associated to the transition between the slow subcritical fracture propagation by stress-corrosion and the dynamic critical propagation around $K_C$. Dynamic fracture and related roughening have already started on regions that are about ten times smaller than the mirror region, the end of which is arbitrarily set by the optical detection limit [2].

The close inspection of fracture surfaces during the genuine stress-corrosion fracture has thus been lacking until very recently. A preliminary study by Wiederhorn et al. [1] using stable crack propagation techniques such as Dual Cantilever Beam (DCB) has shown that fracture roughness in the stress-corrosion regime is on the contrary a decreasing function of the applied stress intensity factor. In this regime where $K_I$ is smaller than $K_C$ and other roughening modes are substantially absent, the crack surfaces do not present any sign of individual damage mechanisms such as multiple crack nucleation or twist bifurcations. The roughness is rather randomly distributed, but it was proven to manifest a self-affine statistical nature [10]. To explain the decrease of roughness with $K_I$, Wiederhorn et al. [1] proposed a model based on the interaction between the slow crack front propagation and the material's heterogeneities. The glass properties were assumed to be elastic down to molecular scales, but possessing a heterogeneous distribution of both elastic moduli and thermal expansion coefficient at some nanometric scale. The deviations of the crack propagation direction from the pure mode I plane are determined by the local mode II stress intensity factor $K_{II}$ induced by the perturbative stress field generated by the material's heterogeneity. The amplitude of this effect is shown to be reduced when the mode I loading is increased in agreement with the experimental observations. However, no quantitative relation is proposed between the characteristic scales of the material's heterogeneity and the amplitude of the surface roughness.

These kinds of investigations are very important to understand both the deep nature of the stress-corrosion mechanisms and their interaction with the nanoscale structural heterogeneity of glassy materials. Stress-corrosion is classically interpreted as a thermally activated and stress enhanced hydrolysis reaction where the main reactant is provided by environmental water molecules [11]. Water molecules were also proved to diffuse into a few nanometer layer below crack surfaces [12], where they may possibly alter the glass structure by inducing local damage or flow [13, 14], or even by inducing compressive stresses and thus a partial shielding of the crack tip stresses [15]. Water molecules coming from the environment could thus play an important role in the local perturbation of the crack propagation direction and thus in the development of the fracture surface roughness.

In this paper, we first present an investigation by Atomic Force Microscopy (AFM) of the separate effects of stress intensity factor, crack tip propagation velocity and relative humidity on the roughness of fracture surfaces obtained by controlled stress-corrosion crack propagation in silica glass Double Cleavage Drilled Compression (DCDC) samples. We then compare the magnitude of the fracture surface roughness of silica glass with two more glasses (sodo-silicate glass and sodium boroaluminosilicate glass), both of which possess a higher degree of heterogeneity, but a different amount of mobile alkali species. We finally discuss our observations in light of their consistency with a modified version of the model from Wiederhorn et al. [1] that allows taking into account the self-affine nature of the fracture surfaces to identify a quantitative link between the heterogeneity and roughness.

# 2 Materials and methods

## 2.1 Fracture propagation

In the present experiments, fractures were initiated and propagated on a DCDC sample using a precision loading apparatus (based on a Microtest load cell produced by Deben, Woolpit, UK, cf. Figure **1**). The DCDC test set-up is particularly convenient for these studies due to its excellent stability and compactness [16, 17, 18, 19].

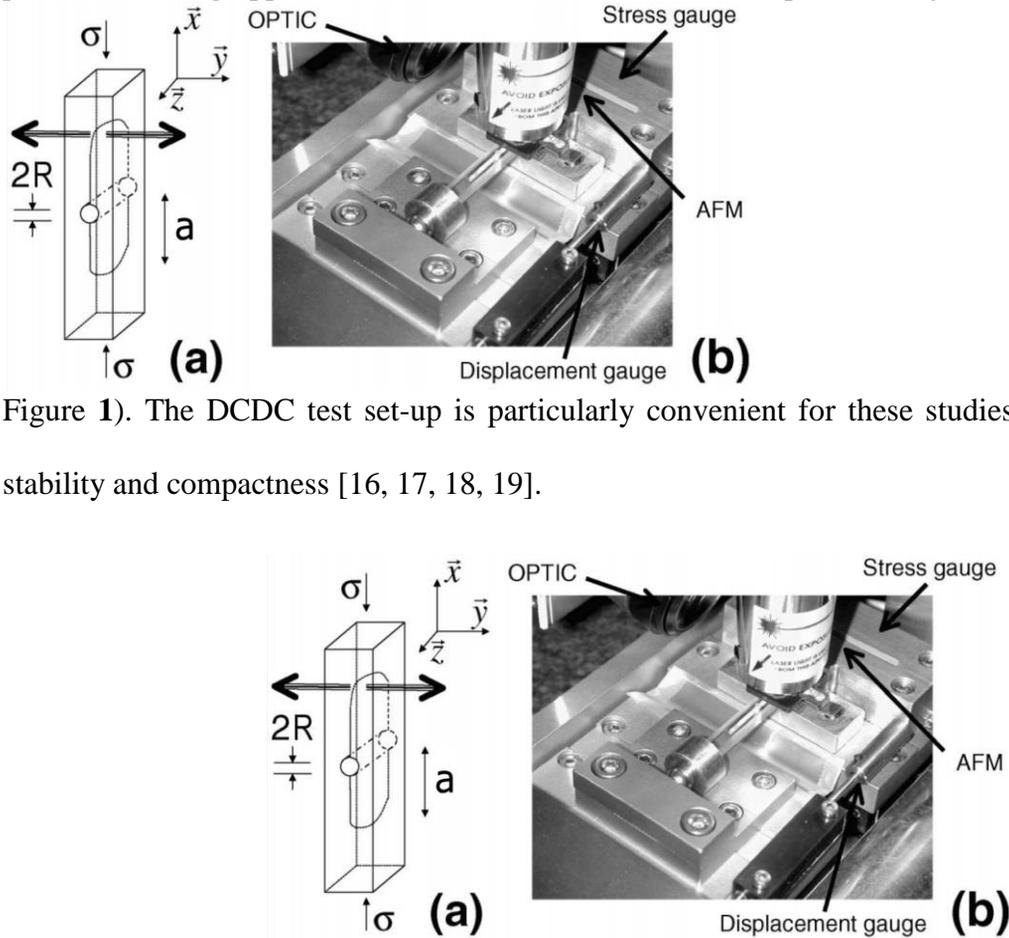

Figure 1: Experimental setup: (a) Sketch of the DCDC geometry; (b) picture of the experiment.

The parallelepipedic DCDC samples of dimensions $2w \times 2t \times 2L$ ($4 \times 4 \times 40$ mm$^3$ $\pm$ 10µm and $5 \times 5 \times 25$ mm$^3$ $\pm$ 10µm) were polished with CeO$_2$ particles to a RMS roughness of 0.5 nm (for an area of $1 \times 1$ µm$^2$) and a hole of radius $R = (500 \pm 10)$ µm was drilled at their center (through the thickness $2t$) to trigger the initiation of the two symmetric fractures of length $a$

(measured from the edge of the hole). As shown in

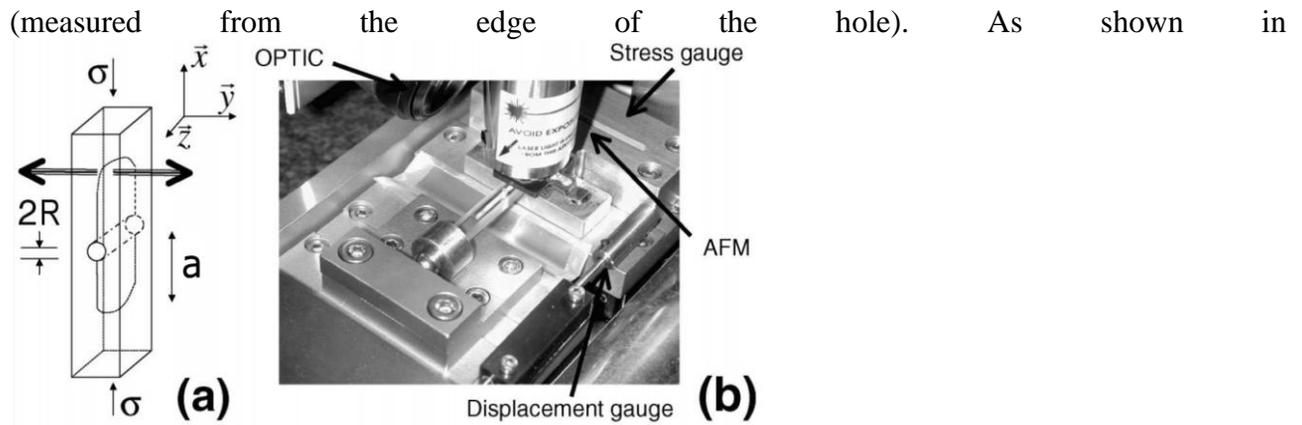

Figure **1**, the $\vec{x}$ and $\vec{z}$ axes correspond respectively to the direction of crack propagation and the direction parallel to the crack front.

Three kinds of oxide glasses were tested: pure fused-silica glass (Corning 7940 and Corning 7980, Corning, USA), sodo-silicate glass (Saint Gobain, France), and sodium boroaluminosilicate glass (Saint Gobain, France). The composition of each glass is given in Table 1. The fractures were propagated in the stress-corrosion regime in a waterproof chamber with a constant temperature $T = (31 \pm 2)°C$ and an atmosphere composed of air and water vapor at a controlled relative humidity level $RH$ between 40% and 100% (the last value corresponding to samples fractured in water).

| Oxide glass | $SiO_2$ | $B_2O_3$ | $Al_2O_3$ | $Na_2O$ |
|---|---|---|---|---|
| Silica | 0.99 | - | - | - |
| Sodo-silicate | 0.75 | - | - | 0.25 |
| Sodium boroaluminosilicate | 0.80 | 0.14 | 0.02 | 0.04 |

Table 1: Composition of the glass specimens in wt fraction.

The stress intensity factor $K_I$, at imposed loading displacement can be evaluated from the value of the measured load $\sigma$, and the optically measured crack length $a$, using Eq. (**1**) [18]:

$$\frac{\sigma\sqrt{\pi R}}{K_I} = \left[c_0 + c_1\frac{w}{R} + c_2\left(\frac{w}{R}\right)^2\right] + \left[c_3 + c_4\frac{w}{R} + c_5\left(\frac{w}{R}\right)^2\right]\frac{a}{R} \qquad (1)$$

with the set of parameters $c_0 = 0.3156$, $c_1 = 0.7350$, $c_2 = 0.0346$, $c_3 = -0.4093$, $c_4 = 0.3794$, and $c_5 = -0.0257$ for $2.5 \leq w/R \leq 5$ and $w < a < L - 2w$.

By coupling optical and atomic force microscopy, the crack propagation velocity $v$ can be measured in a range from $10^{-4}$ m/s down to $10^{-12}$ m/s. For each set of parameters ($K_I$, $v$ and $RH$), the crack is propagated in steady-state conditions in pure opening mode I over a sufficiently long region to allow for a convenient topographic analysis of the fracture surface roughness.

## 2.2 Crack surface measurements by AFM

Once the specimen is fractured into two pieces, the post-mortem fracture surfaces are scanned with the AFM in either tapping mode (Bruker Nanoscope Dimension V, Santa Barbara, CA, USA) or contact mode (NT-MDT, Moscow, Russia). In the fracture surface region corresponding to each set of parameters, a series of AFM images ($512 \times 512$ and $1024 \times 1024$ pixels) is scanned with different sizes ranging from 1 to 5 µm wide squares. The fast scanning axis is chosen to be parallel to the crack front ($\vec{z}$ direction). All the AFM tips used have a nominal radius $R_{tip} \sim 10$ nm (DNP and MPP-11100-10 from Veeco Metrology Inc, Camarillo, CA, USA; Silicon nitride gold coated from ScienTec, Les Ulis, France).

Since the typical RMS values to be measured are about or below 1 nm for our image sizes, it is of particular importance to optimize the imaging parameters and feedback properties in order to keep the instrument noise as close as possible to the lower bound limit (about 0.2 Å). Moreover, since an accurate roughness analysis also relies on the estimation of the lateral correlations of the roughness fluctuations along the fracture surface, an important effort has been spent on insuring an optimal metrological accuracy during the image acquisition. The AFM drifts during the imaging were thus minimized by using a close-loop correction of the lateral motions of the scanning head. The optimization of the scanning parameters was also oriented to a minimization of the wear of the AFM tip, the size of which was periodically monitored in order to avoid progressive bias [20].

## 2.3 Crack surface measurements by AFM

The fracture surfaces of a wide class of materials have long been known to present self-affine statistical properties [21, 22], at least over some finite scaling range. The main consequence is that the RMS roughness $R_q$ evaluated on an image of size $L_0$ according to the definition:

$$R_q(L_0) = \sqrt{\langle y(x,z)^2 \rangle_{x,z}} \qquad (2)$$

(where $y$ is the topographic height as measured from an average reference plane) is an increasing function of the size $L_0$ and it does not provide an absolute estimation of the roughness.

A better description of the roughness properties of a self-affine surface is obtained by evaluating the 1D height-height correlation function over a section $y$ of the topographic image [23], according to:

$$\Delta y(\Delta z) = \sqrt{\langle (y(z+\Delta z) - y(z))^2 \rangle_z} \qquad (3)$$

which essentially estimates the RMS of the height difference between couples of points separated by a distance $\Delta z$. For isotropic self-affine surfaces this quantity is found to scale as a power law of the distance $\Delta z$:

$$\Delta y(\Delta z) = \ell^{1-\zeta} \Delta z^\zeta \qquad (4)$$

where $\zeta$ is the roughness exponent (or Hurst exponent). The second parameter $\ell$ is called the topothesy, and it is a characteristic length-scale of the self-affine surface, i.e. the scale for which the average height difference $\Delta y$ becomes equivalent to the horizontal distance $\Delta z$ between considered couple of points on the fracture surface [23]. The two parameters $\ell$ and $\zeta$ provide a consistent characterization of the self-affine surface roughness, which is independent on the size of the image that was used to evaluate it. Since $\zeta$ is generally between 0 and 1, a self-affine surface will tend to look smooth at scales much larger than the topothesy $\ell$ and very rough at much smaller scales. The RMS roughness $R_q$ can be expressed in terms of $\ell$, $\zeta$ and the size $L_0$ of the image according to:

$$\boldsymbol{R_q(L_0) = A\ell^{1-\zeta}L_0^{\zeta}} \tag{5}$$

where $A$ is a constant close to 1, which is weakly dependent on the image resolution (in terms of pixel size).

By the intrinsic nature of AFM images, the vertical positioning of subsequent scan lines is affected by a small repositioning noise, which has a critical effect on images with very small roughness. The evaluation of our statistical estimators is thus only accurate in the fast scan direction (which is chosen here to be parallel to the crack front). A linear fit was subtracted from each individual scan line to reduce the bias related to both the weak sample tilt and the vertical offset between scan lines. In order to obtain a very robust estimator the average is computed over the whole set of 1D height-height correlation functions obtained for each line.

## 3 Results

### 3.1 Analysis of the scaling regimes

In Figure 2 we represent a typical example of the measured fracture surface topography and of the 1D height-height correlation function estimated according to Eq. (**4**). For all the values of $K_I$ and $RH$, these functions have a similar overall shape as those reported in [24], exhibiting two distinct self-affine regimes at small and large scales, separated by a cutoff length $\xi$ that typically ranges between 20 and 80 nm.

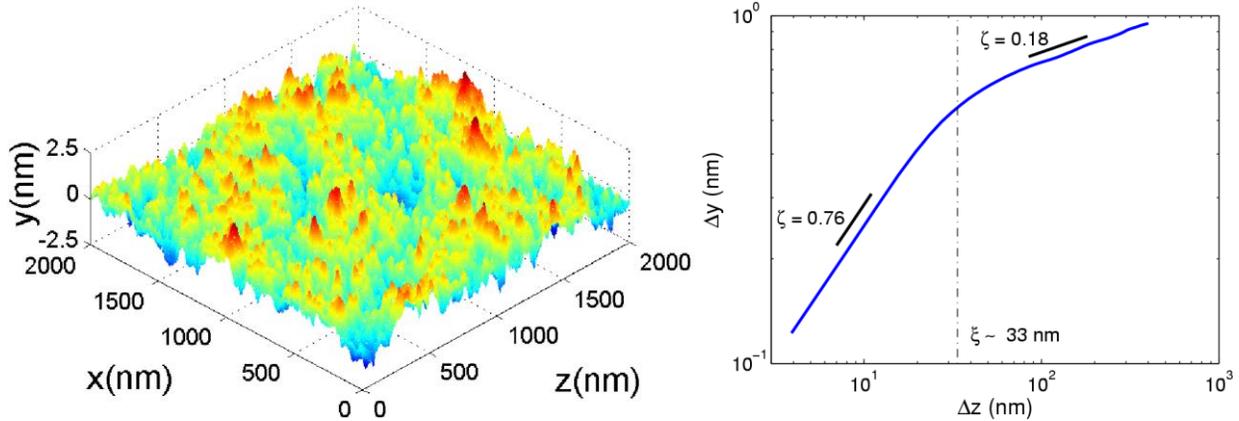

Figure 2: (Left) Experimental fracture surface of fused-silica glass obtained by AFM in tapping mode for a value of stress intensity factor $K_I = (0.34 \pm 0.01)$ MPa·m$^{1/2}$ and relative humidity $RH = (78 \pm 2)\%$. (Right) 1D height-height correlation function corresponding to the same image.

At small scales the roughness exponent $\zeta$ typically ranges between 0.6 and 0.8 (with non systematic dependence on the fracture parameters) and the topothesy $\ell$ ranges between $10^{-15}$ m and $10^{-12}$ m, which is surprisingly small and below the continuum limit of matter. Although in the literature this small-scale regime is frequently attributed to a universal fracture roughness exponent, recent metrological investigations have shown that when measuring extremely smooth surfaces such as glass fracture surfaces by AFM, this regime could simply be generated by the nonlinear smoothing effect related to the finite size of the AFM tip, and it is thus very sensitive on the progressive tip damage [20]. The small-scale regime can thus not be soundly used to infer physical properties of the glass fracture surfaces and will be disregarded here.

We thus decide to focus on the large-scale self-affine regime only, which presents a lower exponent $\zeta = (0.18 \pm 0.10)$, and a topothesy $\ell$ ranging from few Ångströms to nanometers, which is very close to the molecular scales. Since the considered AFM images are larger than 1 μm, the RMS roughness $R_q$ will generally be dominated by this large-scale regime, and the low value of $\zeta$ is indeed consistent with the scaling exponents reported by Wiederhorn et al. [1] for the dependence of $R_q$ on the image size $L_0$ (cf. Eq. (**5**)). We remark that due to the low exponent of the large-scale regime, we cannot exclude a slow convergence towards the logarithmic behavior reported by [24], but the available scaling range is not sufficient to discriminate between the two descriptions.

## 3.2 Effect of loading and environment

The measured values of $R_q$ in silica glass for three different scan sizes are represented in

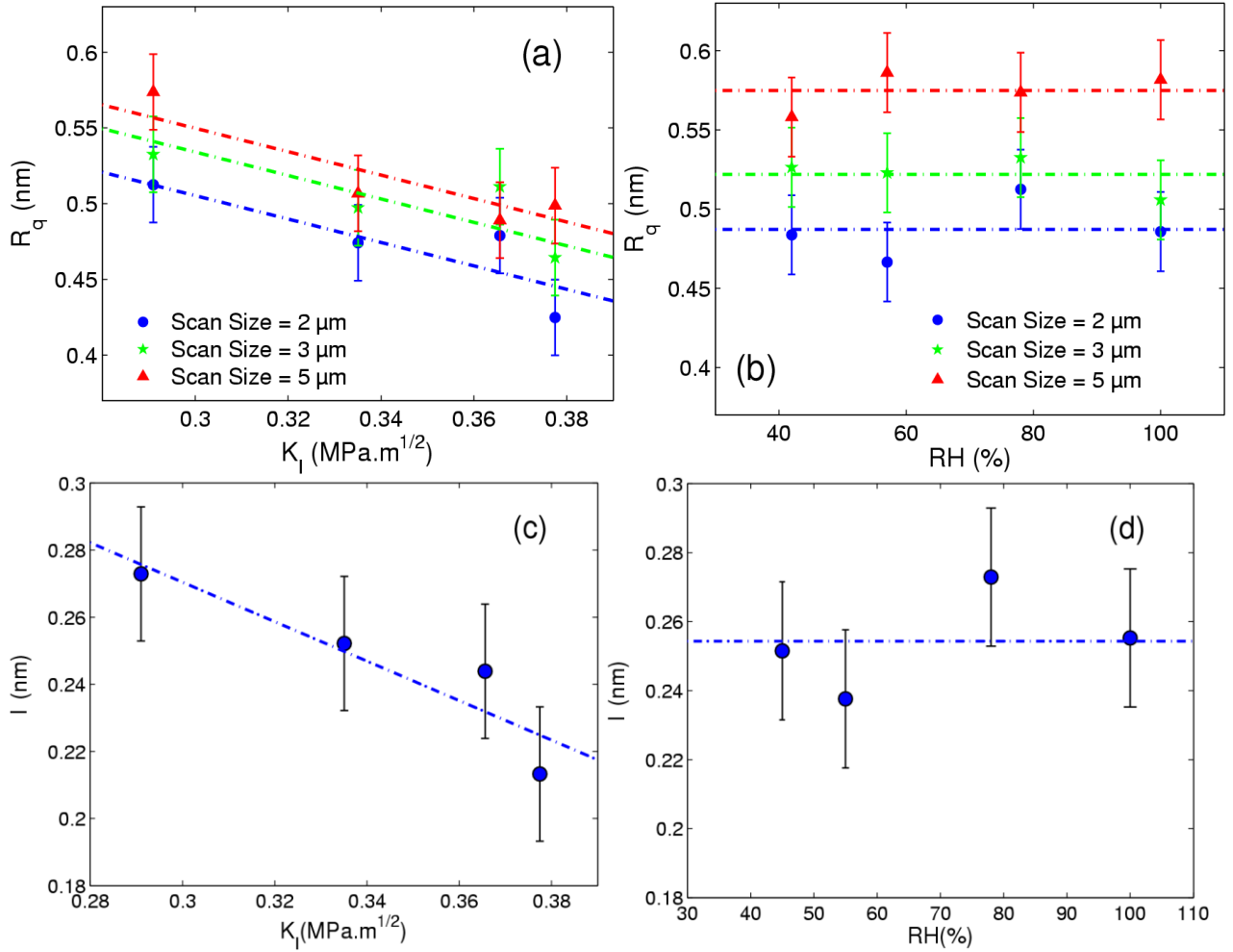

Figure **3**(a) as a function of four different values of $K_I$ at constant $RH = (78 \pm 2)\%$, and in

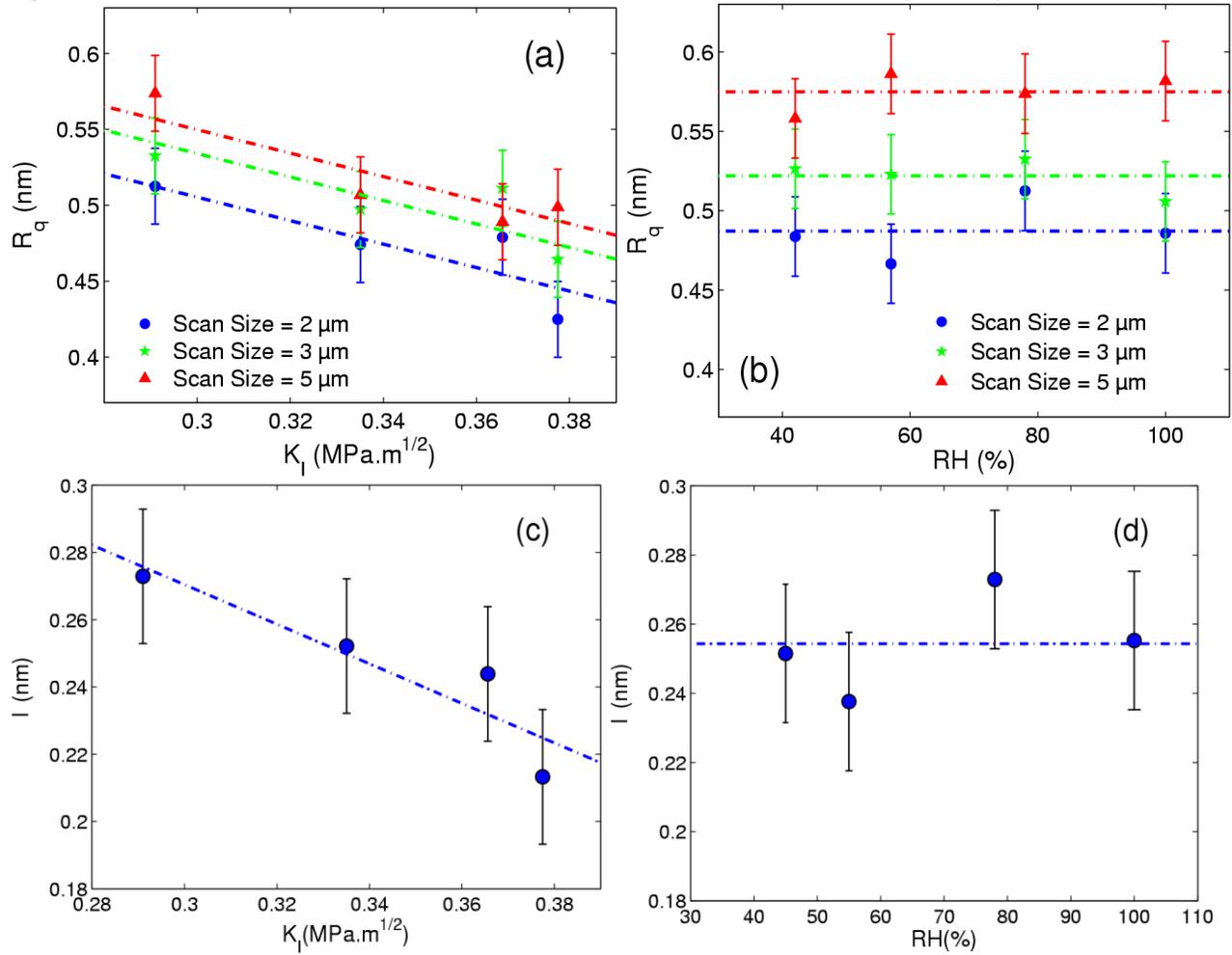

Figure **3**(b) as a function of four different values for the relative humidity $RH$ at constant $K_I = (0.29 \pm 0.01)\ \text{MPa} \cdot \text{m}^{1/2}$ ($RH = 100\%$ corresponds to samples broken in water).

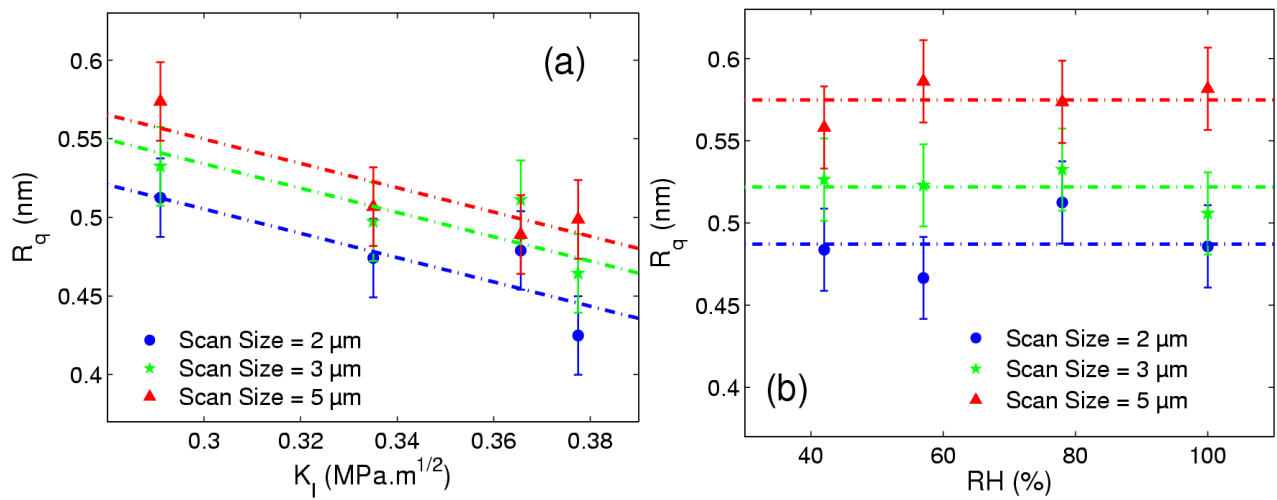

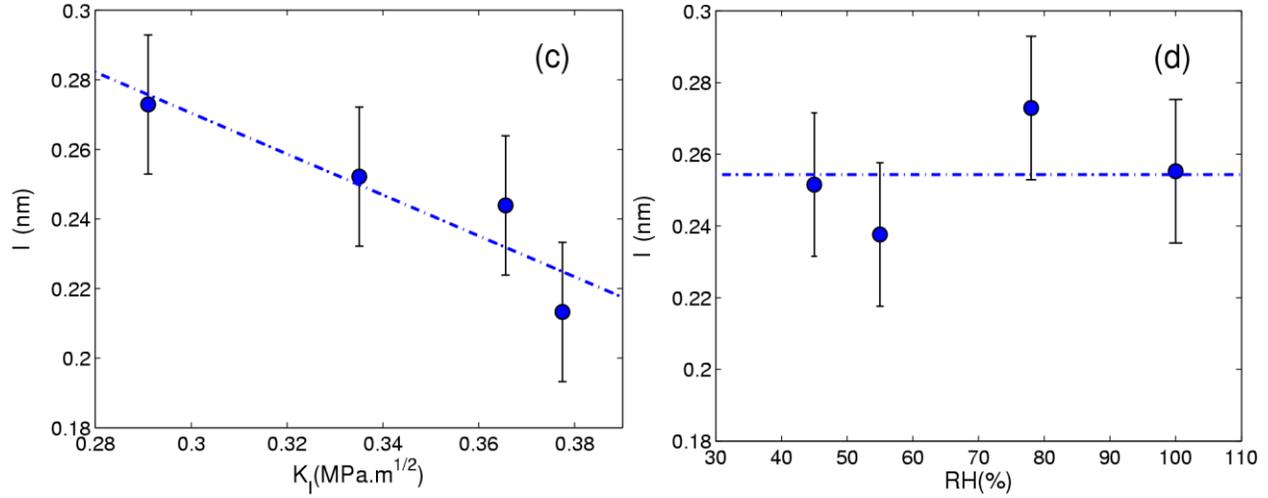

Figure 3: The dependence of the fracture surface roughness on $K_I$ and $RH$ is first reported in (a) and (b) using the traditional $R_q$ estimator for different image sizes, and then in terms of the topothesy $\ell$ in (c) and (d) for a fixed value of the exponent $\zeta$.

For each combination of the fracture parameters, we can combine the height-height correlation functions for the three scan sizes in order to extract a couple of parameters $\zeta$ and $\ell$ that characterize the large-scale regime. When combining all data together, the value of the roughness exponent $\zeta$ does not show any significant dependency on the fracture parameters and glass composition. In order to provide a more robust estimation of the topothesy $\ell$ we thus decided to fix the exponent $\zeta$ to its average value of 0.18. The resulting values of the topothesy $\ell$ are represented

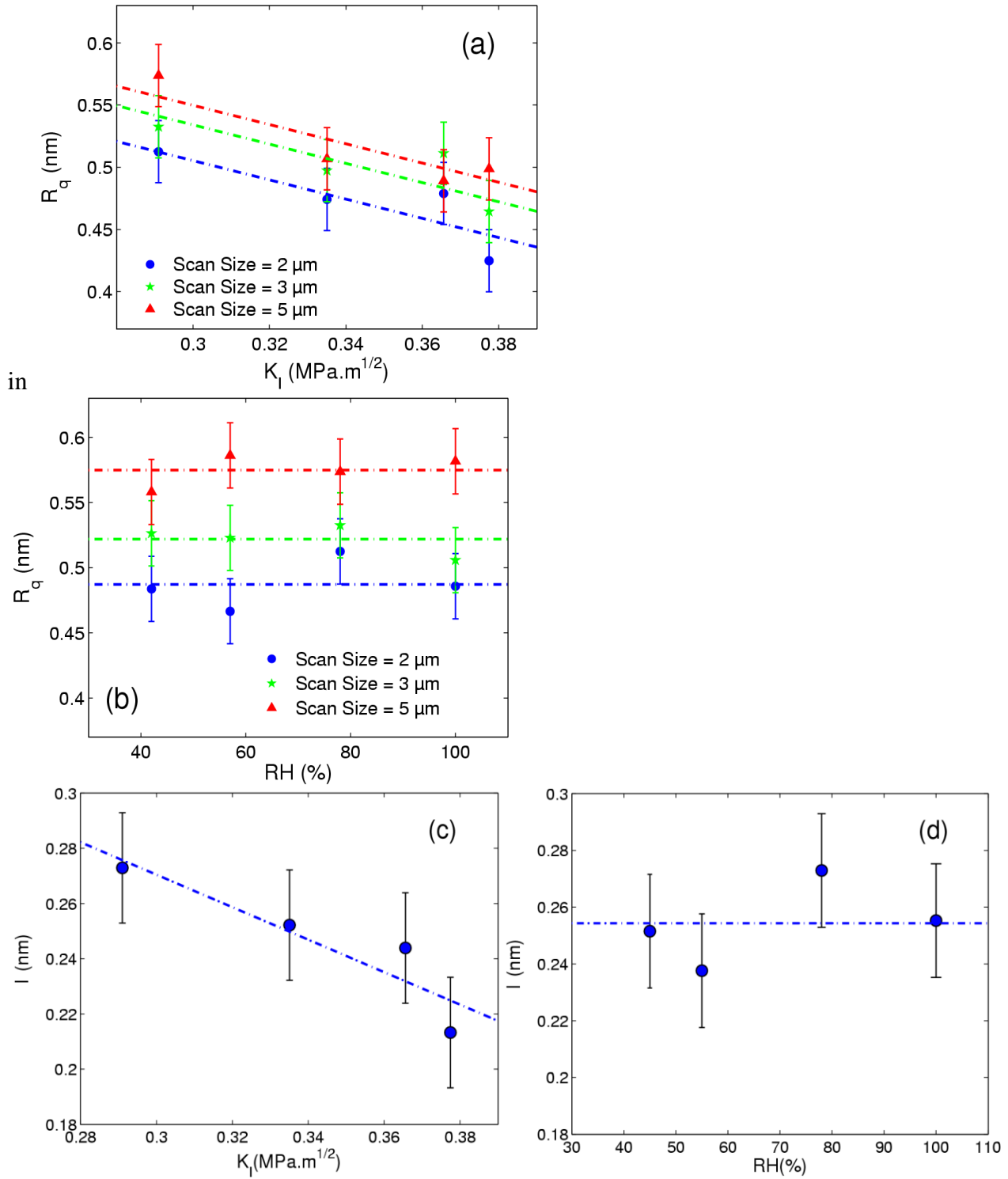

in

Figure **3**(c) and (d).

Our data clearly show that the roughness amplitude, represented by either $R_q$ or $\ell$, has a decreasing trend with $K_I$, which is consistent with the observations by Wiederhorn et al. [1]. On the other hand, the data do not show any significant variation of the roughness over a wide range of relative humidity at constant value of $K_I$, which is an important and original result.

### 3.3 Effect of glass composition

In Figure 4 we compare the measured roughness for the three different glasses (silica glass, sodo-silicate glass and sodium boroaluminosilicate glass) by showing its dependence on the stress intensity factor $K_I$ for a constant relative humidity $RH = (45 \pm 4)\%$.

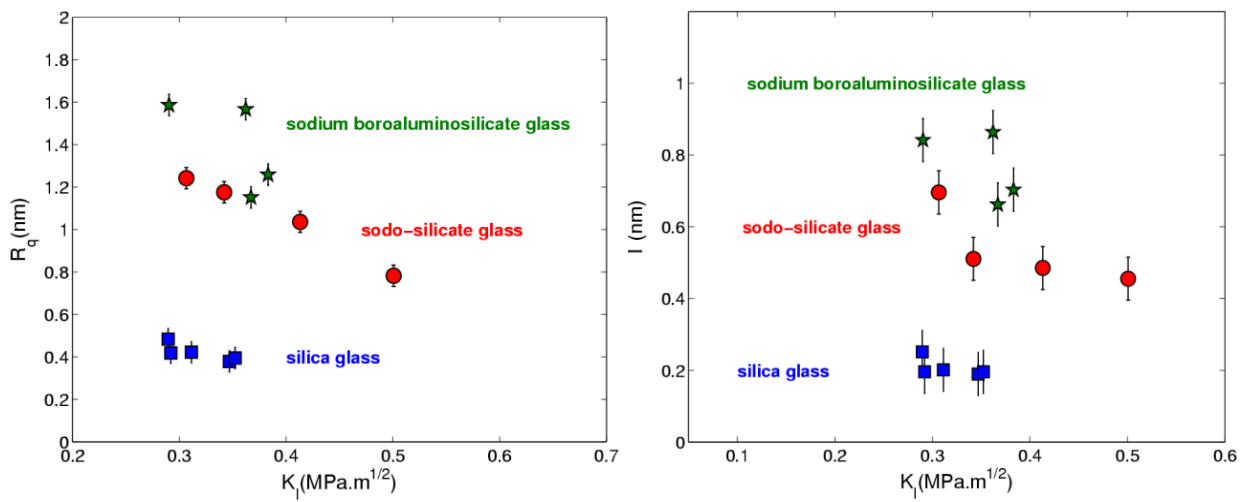

Figure 4: (Left) RMS roughness $\boldsymbol{R_q}$ as a function of the stress intensity factor $\boldsymbol{K_I}$ at $\boldsymbol{RH =}$ $(\boldsymbol{45 \pm 4})\%$ for the three different glass compositions: silica, sodo-silicate, and sodium boroaluminosilicate glass. (Right) Same data analyzed in terms of the topothesy $\boldsymbol{\ell}$ for a constant exponent $\boldsymbol{\zeta}$.

In all the analyzed glasses the roughness, expressed by either $R_q$ or $\ell$, decreases with $K_I$. Our results are consistent with Wiederhorn et al. [1] who argue, as other authors [8, 25], that the roughness in glass results from the interaction of crack propagation with the heterogeneities in the composition and structure of glass. In our results, one can note that the roughness amplitude in sodo-silicate and sodium boroaluminosilicate glass is higher than that of silica. This can be related to a higher degree of heterogeneity in these more complex glasses and will be discussed in more detail later.

## 4 Modeling arguments

In order to interpret our experiments, we present here an extended version of the modeling arguments from Wiederhorn et al. [1], which will be adapted to take into account the self-affine nature of the fracture surfaces and to obtain a more quantitative prediction of the link between the amplitude of the fracture surface roughness and the characteristic length-scales of the material's heterogeneity.

The original model from Wiederhorn et al. [1] is essentially a 2D model based on the interaction of a brittle quasi-static straight propagating fracture with a perturbative random stress field, which is attributed to the nanoscale heterogeneity of the glass properties through the combination of two different effects. The heterogeneity is represented as random inclusions of typical size of a few nanometers, possessing different values of both the elastic modulus and the thermal expansion coefficient. The first component of the perturbative stress field corresponds to the residual thermal stresses originated by the heterogeneity of the thermal expansion coefficient during the slow quench after the glass transition; it is thus supposed to be independent of the external loading $K_I$. The second component of the perturbative stress field is originated form the interaction between the singular crack tip field and the random inclusions due to the mismatch of their elastic modulus; it is thus expected to be proportional to the magnitude of the external loading $K_I$. More precisely, the amplitude of the perturbative stress should be driven by the local stress intensity factor $K_{I,tip}$ at the nanometer scale, since the crack tip region can possibly be shielded by the residual stresses induced by sodium or water penetration in the neighborhood of the crack tip [26, 27, 28].

The two perturbative stress fields interact with the crack propagation by the generation of local mode II stress intensity factor $K_{II}^*$, which can therefore be expressed as:

$$K_{II}^* = A + BK_{I,tip} \qquad (6)$$

where *A* and *B* are two material dependent constants related respectively to the amplitudes of the thermal and elastic heterogeneities. Although a mode I perturbation $K_I^*$ of the same order is also present, it can be neglected in front of the external mode I loading, and the mode II perturbation is thus the relevant one for determining the local mode mixity (defined as $K_{II}/K_I$). According to Gao and Rice [29], the angle of deflection $\theta$ of the crack propagation direction with respect of the straight propagation associated to the external mode I loading, is related to the local mode mixity according to the relation:

$$\theta \sim \frac{K_{II}^*}{K_{I,tip}} \qquad (7)$$

Combining Eqs. (**6**) and (**7**) Wiederhorn et al. [1] obtain the following scaling relation for the deflection angle:

$$\theta \sim \frac{A}{K_{I,tip}} + B \qquad (8)$$

By considering a harmonic perturbation $y(x)$ of wavelength $\lambda$ of the crack path with respect to the straight crack, the characteristic amplitude $y$ of the resulting roughness will be proportional to the angular perturbation since $\theta \sim dy/dx \sim y/\lambda$. The total fracture surface roughness $R_q$ will thus possess the same scaling as $\theta$ in Eq. (**8**) in terms of the stress intensity factor, although the proportionality factor is not provided by the model. The first term represents the roughness generated by thermal stresses and its amplitude is expected to decrease with the stress intensity factor $K_{I,tip}$. The second term represents the contribution of the elastic mismatch and it is independent of the external loading; it should thus constitute a lower bound for the fracture surface roughness, which can be experimentally estimated to the very low values of a few Ångströms [1]. We remark that while this very interesting model provides a sound interpretation of the observed scaling of the surface roughness with $1/K_I$, it cannot predict nor account for the self-affine nature of the fracture surfaces, and it does not provide a clear quantitative relation between the amplitude of the roughness and the relevant length scales corresponding to the material's heterogeneity.

As discussed in section 2.3, the self-affine nature of the fracture surfaces (at least over a finite yet significant scale range) induces a more natural description of the roughness by two intrinsic parameters: the roughness exponent $\zeta$ and the topothesy $\ell$. Any sound modeling for predicting the fracture surface roughness should thus be formulated in terms of these two quantities, instead of the classical RMS roughness $R_q$ that depends on the size $L_0$ of the image to be analyzed. Although many models exist for predicting the roughness exponent $\zeta$ of the self-affine roughness of fracture surfaces based on the statistical physics of the interaction of a propagating crack with several kinds of material's heterogeneity, no clear consensus has emerged on their applicability and on the degree of universality of the proposed roughness exponents [10, 30]. On the other hand, the topothesy $\ell$ can be given a more natural physical interpretation that provides a sound completion to the model of Wiederhorn et al. [1].

Since in all our measurements the exponent $\zeta$ of the significant large-scale self-affine regime (cf. section 2.3) did not show any dependence on the fracture parameters and glass composition, we assumed its average value $\zeta = (0.18 \pm 0.10)$ to be indeed a constant for the three explored silica based glasses. The topothesy $\ell$ thus becomes a robust physical counterpart of the roughness amplitude $R_q(L_0)$ for self-affine surfaces, according to Eq. (**5**). Following the definition (**4**), the topothesy $\ell$ corresponds to the only length-scale where the vertical and horizontal variations of the topography are equivalent, that is, where the average perturbation angle $\theta$ become of the order of 1 (radiant). According to Eq. (**7**), this can be physically related to the physical length-scale where the perturbation $K_{II}^*$ becomes of the same order as the external loading $K_I$ (or $K_{I,tip}$ in case of shielding). By dimensional arguments we can infer that since the topothesy is the only length-scale of a self-affine fracture surface, it should be closely coupled to a relevant length-scale of the material's heterogeneity. Moreover, considering Eq. (**5**) and the fact the exponent $1 - \zeta$ is close to 1, we can expect the topothesy to be affected by $K_I$ in a similar matter as $R_q$.

Although silica based glasses have been shown to be extremely homogenous and amorphous materials down to less than 3 nm scales [31], their structure becomes progressively organized close to the molecular scales, resulting in very large local fluctuations of all their properties, including elastic moduli and thermal properties, which become barely definable. The material's heterogeneity, and the related perturbative stress field discussed above, will thus result to be larger and larger when we reduce the length-scale of observation. If we consider the physics of crack deflection described by the model of Wiederhorn et al. [1] to act at all scales according to the relative fluctuation of the material's properties at the considered scale, than the self-affine nature of the surfaces can be naturally understood. The average crack deflection will be larger and larger at smaller scales due to the increase of the relative amplitude of the fluctuations. The topothesy $\ell$ of the fracture surface roughness should thus naturally be coupled to a natural the length-scale of the material's heterogeneity, which we propose to be the length-scale where the relative fluctuations of its physical properties become close to 1. According to this novel interpretation, it is sound that the range of the measured topothesy of the large-scale regime, between 0.2 and 1 nm, is found to lie between the molecular scale and the typical scale of 3 nm where the physical properties of silica based glasses are reported to be homogeneous [31].

## 5 Discussion and Conclusion

When expressed in terms of $R_q$, our experimental investigation constitutes an extension of the one conduced by Wiederhorn et al. [1] in order to understand the relationship between the fracture surface roughness, the relevant heterogeneity of glass and the nanoscale mechanisms acting at the crack tip in the stress-corrosion regime. Where overlapping, our data are consistent with Wiederhorn's observations and modeling. On the other hand, the independent investigation of the role of humidity and crack propagation velocity provides new insights towards a better understanding of the stress-corrosion mechanisms and the new developments of the modeling provide a more quantitative link with glass heterogeneity based on the self-affine nature of the fracture surfaces.

The topothesy $\ell$ of the measured fracture surfaces behaves in a qualitatively similar way to the RMS roughness $R_q$ measured on an image of constant size of a few microns, but its value can be more soundly related to the material's heterogeneity. For all glasses the roughness is a decreasing function of the external loading $K_I$ at constant relative humidity. However, since the changes in $K_I$ at constant relative humidity also induce an exponential increase of the crack propagation velocity [11], these data alone cannot exclude that the relevant physical parameter affecting roughness be the crack propagation velocity. On the other hand, since our experimental results also show the substantial independence of the fracture surface roughness on the relative humidity at constant $K_I$, we can support more strongly the hypothesis that $K_I$ is the relevant physical parameter for the fracture surface roughness. We recall that in the stress-corrosion regime the crack propagation velocity changes in an almost proportional way to the relative humidity at constant $K_I$ [11], and this does not seem to affect the roughness in an appreciable way.

We will now discuss our results in terms of recent knowledge about the mechanisms of stress-corrosion crack propagation in oxide glasses [32, 28]. On the physico-chemical point of view, stress-corrosion in oxide glasses involves the combination of several kinds of reactions such as hydrolysis of the main network, hydration by molecular water and leaching of the alkali ions by exchange with hydronium. Their kinetic competition and the bulk volume affected in the neighborhood of the crack tip depend in a subtle way on the interplay between the glass composition, the stress intensity factor, the temperature and the local environmental condition at the crack tip. Most of these reactions involve a local expansion of the affected glass region and thus the setup of some additional compressive stresses due to the confinement of the crack tip region into the bulk glass. These are responsible of the shielding of the crack tip stress intensity factor $K_{I,tip}$ that can thus be smaller than the macroscopically set value $K_I$. Moreover, the very strong tensile stresses at the crack tip were frequently argued to induce plastic deformation in a small process zone.

After many years of debate, several very resolved recent experimental observations have led to a global consensus about the relevant scales of these effects in the most common oxide glasses at ambient conditions [28]. In-situ AFM investigations combined with a multi-scale analysis of the whole fracture test samples have allowed to confirm the validity of the perfectly elastic behavior of glass down to 10 nm from the crack tip, which is the limiting resolution for in-situ AFM investigations [33][19]. An important fractographic analysis by [34] has proven that not only the fracture surface roughness is below the nanometer amplitude, but also that the roughness of two opposite crack surfaces has a complementary shape and can be recombined with a resolution better than 0.3 nm, which is close to the limiting resolution of AFM. These combined observations lead to the conclusion that the crack tip plasticity in the stress-corrosion regime should be limited to several molecular distances, in agreement with the predictions from the Dugdale model [28]. Concerning alkali modified glasses such as sodo-silicate glass, Fett et al. [26] have demonstrated that ion-exchange can affect a region of the order of 10 nm, and hypothesized that the induced compressive stresses can be treated as an effective crack tip shielding that is responsible for the presence of a crack propagation threshold in this kind of glasses. Concerning pure silica glass, recent measurements by Lechenault et al. [12] indicate water penetration at the crack tip over 6-7 nm, although the molecular form of this water remains debated. Recent investigations by Wiederhorn et al. [35] report some first evidences of a mechanical effect induced by the crack tip water penetration at higher temperature, and argue that these can be attributed to the expansion related to some degree of bulk hydrolysis. Even more recently, Barlet et al. [36] have shown that the intrinsic structure at the nanometric mesoscale of the glasses deeply affects the stress-corrosions cracking behavior.

The overall emerging picture has motivated our choice to use the modeling from Wiederhorn et al. [1] as a starting point for our modeling strategy of the origins of the fracture surface roughness during the stress-corrosion crack propagation in oxide glasses. In this model the mechanical properties of glass are assumed to be essentially elastic at the nanometric scale where the material's heterogeneity (of the elastic moduli and thermal expansion coefficients) is supposed to become relevant. Our development of this model, based on the self-affine nature of the fracture surfaces, allows to link the relevant scale of the material's heterogeneity with the measured topothesy. For silica glass the value of $\ell$ ranges from 0.21 to 0.27 nm, i.e. is comprised between the interatomic distance and structural rings size [9, 37]. If we compare the topothesy values for different glasses at a fixed condition $K_I = (0.3 \pm 0.02)$ MPa·m$^{1/2}$ and $RH = (45 \pm 4)\%$, we observe that $\ell$ is quite larger for the silicate glasses with more complex composition and structure ($\ell_{Boro} = (0.85 \pm 0.05)$ nm, $\ell_{Soda} = (0.70 \pm 0.05)$ nm, to be compared to $\ell_{SiO2} = (0.27 \pm 0.02)$ nm. We remark that the decrease of the fracture surface roughness with $K_I$ in the sub-critical propagation regime leads to the existence of a minimum roughness in the mirror region separating the sub-critical regime from the well known roughening associated to the increase in $K_I$ in the critical propagation regime. The value of the minimum roughness, expressed in terms of topothesy, provides a the characteristic length-scale of the heterogeneity in the local elastic modulus. According to our data and the ones from Wiederhorn et al. [1] for silica glass this limiting value is very close to the molecular size, in good agreement with the estimates of excellent homogeneity of silica glass obtained by molecular dynamics simulations [31]. The main difference between the alkali glasses and silica is in the contribution of the thermal quench term, which induces a stronger roughening at low $K_I$, which is consistent with the tenfold larger value of their thermal expansion coefficient [38]. We remark that the present modeling does not account for the detailed structure of the glass network, and in particular for the level of network depolymerization that has recently been shown to be very promising for understanding stress-corrosion cracking [36,39,40], and should be investigated in future developments.

The present results provide a further consistent observation in favor of the small subnanometric mesocale of the process zone in oxide glasses. In should be noted that these statement is contradictory with the analysis of the same kind of fracture surfaces by Bonamy et al. [22], who interpreted the cutoff length-scale $\xi$ in the 1D height-height correlation functions (cf. Figure 2) as the signature of the size of a plastic process zone of a few tens of nanometers. However, as stated in section 2.3, the measurement of $\xi$ is strongly biased by the progressive wear of the AFM tip and can be shown to reduce progressively when paying more and more attention to avoid the wear of the AFM tips, which have an initial nominal radius of 10 nm [20]. Moreover, the topothesy of the small-scale regime would have non physical values as small as $10^{-15}$ m. We thus confirm that only the large-scale regime seems to carry relevant physical information on the micromechanics of crack propagation. As a final remark, the lack of a detectable dependence on the relative humidity of the roughness of the fracture surfaces in pure silica glass, also provides an important constraint for future modeling about the effect of water penetration on the stress-corrosion mechanisms. Relative humidity is known to affect the pressure and chemical potential of the water condensate at the crack tip, which in turn determines the rate of the stress-corrosion reactions [41, 32, 28]. It can thus in principle also affect the diffusion of water into the glass, that can have two main effects: inducing local compressive stresses and/or weaken the local mechanical properties. Since both phenomena would imply the roughening through either a shielding of the local stress intensity factor or an increase of the local heterogeneity, the experimental accuracy of our measurement allows to limit the variations of the shielding effect to $\Delta K_I \sim 0.1$ MPa·m$^{1/2}$ and to show that the characteristic length-scale of heterogeneity remains of the order of few molecular sizes over the broad humidity variation between 40% and saturation.

## Acknowledgement

We thank L. Ponson, D. Bonamy, D. Vandembroucq and S.M. Wiederhorn for fruitful discussions.